# Parallel Computing Environments and Methods for Power Distribution System Simulation

Ning Lu, Z. Todd Taylor, Dave P. Chassin, Ross Guttromson, and Scott Studham

*Abstract* — **The development of cost-effective high-performance parallel computing on multi-processor supercomputers makes it attractive to port excessively time consuming simulation software from personal computers (PC) to super computes. The power distribution system simulator (PDSS) takes a bottom-up approach and simulates load at the appliance level, where detailed thermal models for appliances are used. This approach works well for a small power distribution system consisting of a few thousand appliances. When the number of appliances increases, the simulation uses up the PC memory and its runtime increases to a point where the approach is no longer feasible to model a practical large power distribution system.**

**This paper presents an effort made to port a PC-based power distribution system simulator to a 128-processor shared-memory supercomputer. The paper offers an overview of the parallel computing environment and a description of the modification made to the PDSS model. The performance of the PDSS running on a standalone PC and on the supercomputer is compared. Future research direction of utilizing parallel computing in the power distribution system simulation is also addressed.**

*Index Terms—***parallel computation, load modeling, message passing interface (MPI), multi-processor supercomputer.**

## I. INTRODUCTION

PHYSICALLY-BASED modeling approaches [1][2] have been widely used to simulation power distribution system loads, because they are able to predict the individual load dynamic response to ambient temperature variations, random customer energy consumption, as well as the electricity market prices. Using this approach, the Energy Science and Technology Division (ESTD) within the Pacific Northwest National Laboratory (PNNL) has created a prototype ultra complex power system simulator [3][4][5] (called Power Distribution System Simulator or PDSS) under a Laboratory Directed Research and Development project titled the Energy System Transformation Initiative. The key feature of this computer program is its ability to accurately predict load shapes of various household appliances such as building heating ventilation and air conditioning (HVAC) systems, refrigerators, lighting, washers and dryers, dishwashers and ranges. By solving the power at each individual load using first principle calculations, this approach enables one to simulate the price responsive load control technology under various market structures. By aggregating individual loads at the feeder level, one can simulate the aggregated response of these loads with reasonable accuracy. This approach works well for a power distribution system of a few thousand houses. When the number of houses increases to around 10,000, its runtime increases rapidly and the simulation uses up the PC memory, making it infeasible to model a practical large power distribution system.

One option to reduce the simulation time and conform to the memory requirement is to conduct the simulation in multi-processor supercomputers. The development of cost-effective high-performance parallel computing on multi-processor supercomputers makes it attractive to port excessively time consuming simulation software from PCs to supercomputers. PNNL acquired an SGI Altix 3000 128 CDM SMP superconomputer for general purpose high performance computing applications. The system runs a single Linux operating system over 128 Intel Itanium 2 processors running at 1.5 GHz. In addition, the system has 256 GB of RAM and ¼ TB of disk space. An effort is made to port the PC-based PDSS to the Altix supercomputer. In this paper, the technique used to modify the PDSS for parallel computation is presented. The results indicate that the parallel computing approach works well in physically-based distribution system simulations. The memory requirement is met and the runtime is significantly shortened.

## II. THE MODELING APPROACH OF PARALLEL COMPUTATION

### A. The modeling approach of the PDSS

The Power Distribution System Simulator developed by PNNL takes a bottom-up approach, in which detailed physically-based models of each type of appliance are developed. As shown in Fig.1, residential appliances are categorized into those that are thermostatically controlled and those that are non-thermostatically controlled. There are three

This work is supported by the Pacific Northwest National Laboratory, operated for the U.S. Department of Energy by Battelle Memorial Institute under contract DE-AC06-76RL01830.

Ning Lu, Z. Todd Taylor, Dave P. Chassin, Ross Guttromson, and Scott Studham are with Pacific Northwest National Laboratory, P.O. Box 999, MSIN: K5-20, Richland, WA - 99352, USA. (e-mails: ning.lu@pnl.gov, todd.taylor@pnl.gov, david.chassin@pnl.gov, ross.guttromson@pnl.gov, scott.studham@pnl.gov.)

types of thermostatically controlled appliances (TCAs) allowed in the model: heating ventilation and air conditioning (HVAC) systems, electric water heaters, and refrigerators. Non-thermostatically controlled appliances include dish washers, washers/dryers, and the like.

There are six appliance modules in PDSS, as shown in Fig. 2. $P_L$ is the power at the feeder head, $P_H$ is the power output of a household, and $B$ is the energy market price. The inputs of each household load are temperature data, setpoint setting sequence, customer consumption probabilities, electricity market prices, and time steps. During the initialization, PDSS reads the data and allocates memory for each house. PDSS then calculates the power output for each appliance in a household and then aggregates its energy consumptions at each time step. For a non-TCA, PDSS determines its on/off status based on its likelihood of on/off. For a TCA, PDSS calculates the inflection points, where the appliance will turn-on or turn-off, based on the current ambient temperature and its temperature setpoint. As shown in Fig. 2, the calculation of each household load is a highly parallel process and is suitable for parallel computation.

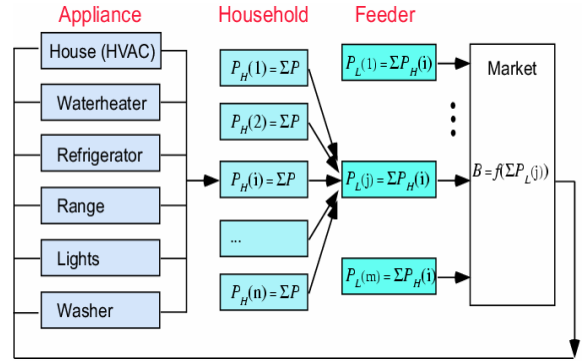

Fig. 2: The block diagram of PDSS

### B. The parallel processing modeling approach

There are two approaches to apply parallel computation for the PDSS based on its software structure. The first approach is the shared-memory approach; the second one is the message passing interface (MPI) approach.

#### 1) The shared-memory approach

The shared-memory [6][7] approach provides tasks with a common asynchronous read/write-shared-address space, where access is controlled by lock and semaphore mechanisms.

As shown in Fig. 3, all the house data are put into the shared-memory. The calculation of individual house loads is evenly distributed among all the processors. The results obtained are then written back to the shared-memory. The advantages of this approach are:

- The runtime can be shortened significantly. Communication between processors is minimized because of the shared-memory.
- The calculation is efficient because the houses are evenly distributed to processors.
- Synchronization is done at the end of each time step.

However, the approach requires significant coding efforts to implement. Because the standalone version of the PDSS is not developed for parallel computation, new codes need to be written to modify the current PDSS codes for memory allocation. The shared-memory will be used to store the house parameters, and any updates after function calls need to write back to the shared memory.

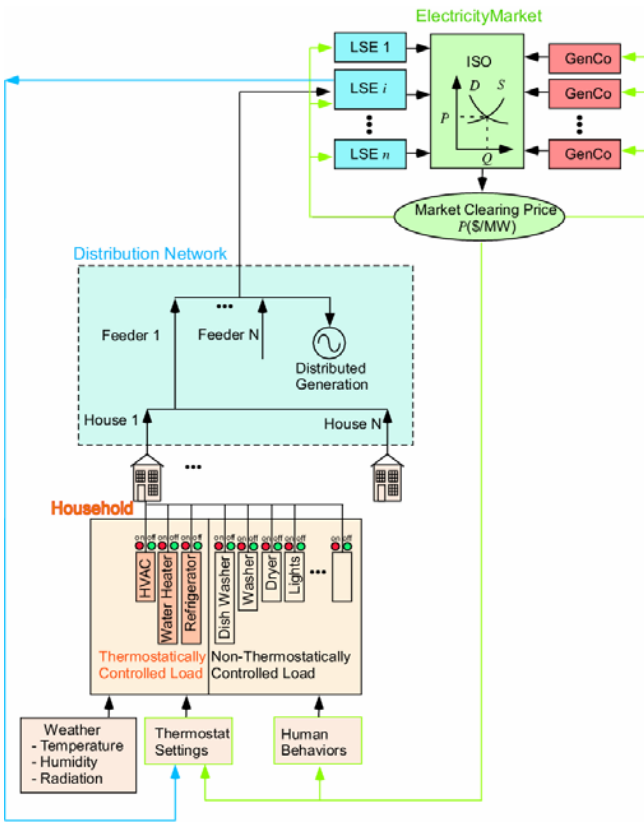

Fig. 1: A bottom-up approach

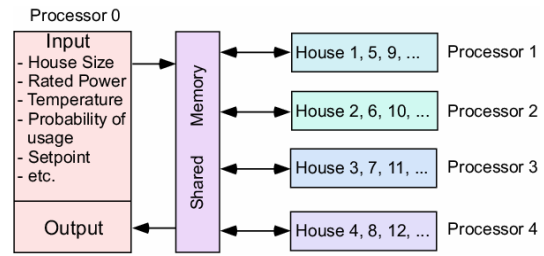

Fig. 3: An example of the shared-memory approach

In addition, in power distribution systems, the residential load will mix with other load at the distribution substation feeder head as shown in Fig. 1. Each feeder made of purely residential load will normally consist of a few hundred to a few thousand households at maximum. It takes a standalone PDSS (refer to $t^1$ in Table I) around 0.71 second CPU time to simulate a 1000-household load over a 100-hour period at 1 hour interval. To split the work among four processors (refer to $t^4$ in Table I), one can shorten the simulation time to 0.25 second (*elapsed* time). The runtime reduction is not significant for parallel the simulation of houses if the number of the houses at a feeder does not exceed 1000.

Furthermore, the control strategies, the weather tapes, as well as price inputs either vary feeder by feeder or by control groups. Thus, to account for the variations, it is easier to parallelize the calculation at the feeder level and pass around the aggregated load information. This leads to the second approach, the MPI approach.

Table I: Simulation Time (Elapsed Seconds)

| Number of Households | 1000 | 5000 | 10000 |
|---|---|---|---|
| $t^1$ (s) | 0.71 | 2.31 | 4.54 |
| $t^4$ (s) | 0.25 | 0.55 | 0.91 |
| $\Delta t = t^1 - t^4$ (s) | 0.46 | 1.76 | 3.63 |

### 2) The MPI approach

As shown in the five-processor example (Fig. 4), the MPI approach [8] is to parallelize the running of PDSS instead of the houses inside PDSS. Let Processor 1 to 4 each run a copy of PDSS, with each PDSS simulating a distribution feeder. Let Processor 0 collect the aggregated outputs and process them. Processor 0 can broadcast general information such as the current market price to Processor 1 – 4, which will then respond accordingly. The key to making this approach efficient is to keep the communication between processors at minimum. To do so, one needs to specify the data that needs to be sent back and forth at the end of each time interval. Currently, we only collect the aggregated power output from each feeder and send the feeders the price data whenever the price changes.

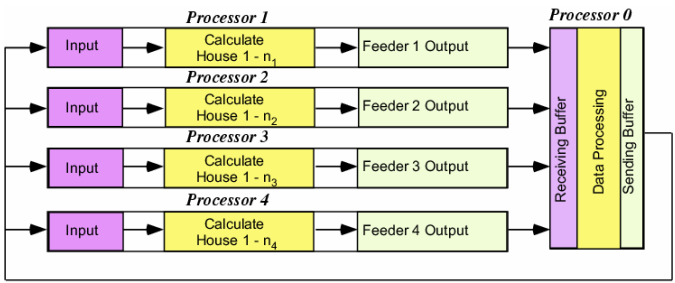

Fig. 4: An example of the MPI approach.

The advantages of this approach are:

- Each PDSS has its own set of input data stored in the local memory of each processor. Therefore, one can run different load control strategies under different weather conditions.

- Minimum coding effort and higher portability. MPI is designed for high performance on both massively parallel machines and clusters, and is a standard for message passing in the parallel computing paradigm. One can call MPI functions in C++ programs to parallel the running of those programs without changing the structure of the programs. The code was tested on a four-processor computer, and then was run on the 128-processor Altix. No additional codes needed to be written.

The disadvantage of this approach is that the total runtime is determined by the feeder serving the most houses. For example, if there are four feeders containing 100, 500, 1000, and 1500 houses, the other processors will have to wait at each aggregation time step until the one simulating the 1500 houses finishes its calculation. Another disadvantage is that each PDSS may run at its own time step. If there are some exogenous changes, such as price, one needs to synchronize each PDSS at each data sending and receiving point.

## III. THE MODELING ENVIRONMENT

In our project, we used MPI on an SGI Altix 3700 super-cluster. The computer consists of 128 Itanium 2 1.5 gigahertz processors, and ¼ terabytes of globally addressable system memory. Communication takes place over the NUMAlink interconnect fabric, which provides latency as small as 50 nanoseconds and a bandwidth of 3.2 gigabytes/second. Altix1 runs a single modified Red Hat advanced server Linux operating system. This presents the image that all the processors in the system are available, as if it were a workstation with 128 processors.

Altix1 recommends that any job taking more than 30 seconds of CPU time use the batch system, for reasons of fairness as well as system efficiency. Altix1 uses the LSF scheduler from Platform Computing to manage resources and schedule jobs on the system. Gold allocation system is used for job submission. LSF is a program that attempts to balance the resource utilization on one or more computers among competing users and their processes. The goal is to give many users' processes a "fair share" of CPU, memory and other resources [9]. LSF works with a set of job queues. Many different queues can be defined, each with its own criteria for job resource allocation.

## IV. PERFORMANCE RESULTS

To determine the best approach to parallelize the running of PDSS, we first ran PDSS on a single processor PC and then conducted three experiments on the Altix supercomputer. All cases are run for 100 hours at 1-hour interval.

### Case 1: Running on a standalone single processor PC

Using a single processor PC, which has a Pentium 4 CPU (2.8 GHz) and 512 MB RAM, we evaluated the runtime of PDSS for a distribution power grid consisting of $10^4$, $10 \times 10^4$, $20\times10^4$, and $50\times10^4$ houses. The results are listed in Table II.

Table II: The Runtime of PDSS (PC case)

| Number of Houses | 10000 | 100000 | 200000 | 500000 |
|---|---|---|---|---|
| Elapsed Time (s) | 11.4790 | N/A | N/A | N/A |

When the number of houses is greater than $10\times10^4$, the PC runs out of memory and PDSS fails to complete the simulation. Therefore, $10\times10^4$ is a threshold when the PC resource has been depleted.

*Case 2: Running on a single processor of Altix1*

The base case is to run $10^4$, $10\times10^4$, $20\times10^4$, $50\times10^4$ houses on a single processor. Assuming that each feed is composed of $10^4$ houses, these cases simulate 1, 10, 20, and 50 feeders, respectively. The CPU time $t_{cpu}$ is the total time spent by the CPU on behave of PDSS and the wall clock time $t_w$ is the true runtime (elapsed time) of the program. The resource usages are shown in Table III.

There are three observations:

- As expected, the runtime is significantly shortened compared with the runtime on PC.

- As expected, both CPU time and elapsed time go up almost linearly with the number of houses. We see from the memory report that the simulation takes enormous physical memory and swap memory. Therefore, a PC's resources can be easily depleted.

- The system efficiency is higher when running more houses. This is because the overhead time is relatively fixed. Therefore, the overhead time takes less share in the total time consumed, when the core program runs longer. Note that when the number of houses is greater than $20\times10^4$, the efficiency increase reaches its peak.

Table III: The Resource Usage Summary (single processor)

| No. of Processors | 1 | 1 | 1 | 1 |
|---|---|---|---|---|
| No. of Houses | $10^4$ | $10\times10^4$ | $20\times10^4$ | $50\times10^4$ |
| CPU Time (s) | 0.42 | 33.64 | 79.17 | 198 |
| Wall Clock Time (s) | 5.28 | 42.98 | 83.03 | 208.60 |
| Efficiency ($t_{cpu} / t_w$) | 8% | 78.3% | 95.4% | 94.5% |
| Max Memory (MB) | 5 | 152 | 284 | 680 |
| Max Swap (MB) | 10 | 4049 | 4181 | 4577 |

*Case 3: Running on multiple processors ($10^4$ houses per processor)*

With each processor allocated one feeder of $10^4$ houses, we then run PDSS on 1, 10, 20, and 50 processors to simulate one-feeder ($10^4$ houses), 10-feeder ($10\times10^4$ houses), 20-feeder ($20\times10^4$ houses), and 50-feeder ($50\times10^4$ houses) cases. We have the feeders running completely in parallel and the communication between feeders happens at the end of the simulation (at 100th hr). The resource usage summary is shown in Table IV.

There are several observations based on the results:

- Compared with the non-parallel case shown in Table III, the runtime is shortened in proportion to the number of processors used. The runtime of the *one-processor-per-feeder* case is almost equal to that of single processor simulating one feeder case. This is as expected because we have limited the number of communications between feeders, which makes the overhead time spent on communication negligible.

- The memory usage is regular. Because each processor is running a simulation of $10^4$ houses, the memory occupation is distributed to each processor. Therefore, it is not necessary for Altix1 to arrange Swap to meet the needs of house parameter storage.

- The efficiency is also similar to that of the single processor case.

Table IV: The Resource Usage Summary (one feeder per processor)

| No. of Processors | 1 | 10 | 20 | 50 |
|---|---|---|---|---|
| No. of Houses | $10^4$ | $10\times10^4$ | $20\times10^4$ | $50\times10^4$ |
| Total CPU Time (s) | 0.42 | 0.42 | 0.44 | 0.44 |
| Wall Clock Time (s) | 5.28 | 5.38 | 5.41 | 5.54 |
| Efficiency ($t_{cpu} / t_w$) | 8% | 7.8% | 7.4% | 8% |
| Max Memory (MB) | 5 | 5 | 5 | 5 |
| Max Swap (MB) | 10 | 10 | 10 | 10 |

From the results, one can reach the conclusion that by simply running the simulations in parallel, we can shorten the simulation time significantly, depending on how many processors one has.

*Case 4: Running on multi- processors ($10^4$ houses in total)*

To study the optimal number of houses to partition to each processor, a total number of 10,000 houses is divided among 10, 20, and 50 processors. As shown in Table V, the runtime is first shortened when dividing the 10,000 houses to 10 processors. However, when dividing the 10,000 houses to 20 or 50 processors, the runtime starts to increase. This is because the overhead time spent on communication and on partition jobs starts to take a greater share in the total program runtime (the wall clock time). Therefore, parallel running very small numbers (200 or 250) of houses will result in an increase of runtime. Thus, the number of houses simulated by each processor should be more than 1000 to make good use of parallel computation.

Table V: The Resource Usage Summary (one feeder per processor)

| No. of Processors | 1 | 10 | 20 | 50 |
|---|---|---|---|---|
| No. of Houses on Each Processor | 10,000 | 1,000 | 250 | 200 |
| Total No. of Houses | $10^4$ | $10^4$ | $10^4$ | $10^4$ |
| Total CPU Time (s) | 0.42 | 0.54 | 0.47 | 0.42 |
| Wall Clock Time (s) | 5.28 | 0.7 | 1.39 | 1.38 |

*Case 5: Running multi- processes on a single processor*

If the number of feeders exceeds the number of processors, one needs to run multi-process on each processor. To study the impact of running multi-process on each processor, a total number of 10,000 houses is divided to 1, 10, 20, and 50 processes and has been run on the 4 processors of Altix1. As shown in Table VI, we observed that the runtime is similar to or longer than the single processor case. Therefore, one may need to combine the simulation of the many small feeders to a few major feeders whenever it is possible.

Table VI: The Resource Usage Summary (4-processor case)

| No. of Processes | 1 | 10 | 20 | 50 |
|---|---|---|---|---|
| No. of Processors | 4 | 4 | 4 | 4 |
| No. of Houses | $10^4$ | $10 \times 10^4$ | $20 \times 10^4$ | $50 \times 10^4$ |
| Total CPU Time (s) | 4.159 | 42.462 | 88.352 | 221.022 |
| Wall Clock Time (s) | 5.210 | 43.550 | 89.340 | 222.100 |

To summarize, we conclude
- The runtime of running the parallel version of PDSS is shortened proportional to the processors used.
- To make the parallel computation worthwhile, the number of houses running on each processor is better above 1000 houses.
- For cases having less than 20000 houses, a single processor will be competent for the simulation.
- Running multi-process on a single processor is not going to save runtime.

## V. Conclusion

This paper presents the results obtained by an effort made to port a serial software package used for power distribution system simulation to a parallel computation environment. The physically-based residential load simulation is especially amiable to parallel computation because there is minimal interaction between each household load. To simulate each individual load with a detailed thermal model, large memory storage is needed. The solving of distributed differential equations also takes significant computer time. To port the simulation to supercomputers, the program runtime is shortened significantly. With 256 GB RAM and ¼ GB disk space, one can simulate the power consumption of millions of households within minutes.

Future research direction will be to fully utilize the MPI and feed price information to each feeder and study the interactive behavior among loads that reside in different distribution feeders. It is also promising to port power flow calculation software packages to the supercomputer and link the distribution system simulation together with the transmission simulation to study the responsive load impact to the overall power systems. Because the distribution simulation and the transmission simulation can be done in a pipelining manner, it can fully utilize the parallel computer resources and extend the simulation ability to the whole power network.

## Acknowledgment

The authors would like to thank their colleague Kenneth P Schmidt for the technical support and suggestions on setting up the experiments.

## References

[1] M. L. Chan, E. N. Marsh, J. Y. Yoon, G. B. Ackerman, and N. Stoughton, "Simulation-based Load Synthesis Methodology for Evaluating Load-management Programs," *IEEE Trans. on Power Apparatus and Systems*, vol. PAS-100, pp. 1771-1778, Apr. 1981.

[2] A. Molina, A. Gabaldon, J. A. Fuentes, and C. Alvarez, "Implementation and Assessment of Physically Based Electrical Load Models: Application to Direct Load Control Residential Programmes," *Generation, Transmission and Distribution, IEE Proceedings*, vol. 150, pp. 61-66, 2003.

[3] R. T. Guttromson, D. P. Chassin, and S. E. Widergren, "Residential Energy Resource Models for Distribution Feeder Simulation," *Proc. of 2003 IEEE PES General Meeting*, Toronto, Canada, pp. 108-113, 2003.

[4] N. Lu and D. P. Chassin, "A State Queueing Model of Thermostatically Controlled Appliances," *IEEE Trans. on Power Systems*, vol. 19, no.3, pp. 1666-1673, Aug. 2004.

[5] N. Lu, D. P. Chassin, and S. E. Widergren, "Modeling Uncertainties in Aggregated Thermostatically Controlled Loads Using a State Queueing Model," submitted to *IEEE Trans. on Power System*, 2004.

[6] G. Fadlallah, M. Lavoie, and L.-A. Dessaint, "Parallel Computing Environments and Methods," *Proc. of 2000 International Conference on Parallel Computing in Electrical Engineering*, pp. 2-7, Aug. 2000.

[7] Michael J. Quinn, *Parallel Programming in C with MPI and OpenMP*. New York: McGraw-Hill, 2004.

[8] Available at: http://www-unix.mcs.anl.gov/mpi/

[9] Available at: http://www-cdf.fnal.gov/offline/runii/fcdfsgi2/

**Ning Lu** (M'98) received her B.S.E.E. from Harbin Institute of Technology, Harbin, China, in 1993, and her M.S. and Ph.D. degrees in electric power engineering from Rensselaer Polytechnic Institute, Troy, New York, in 1999 and 2002, respectively. Her research interests are in modeling and analyzing deregulated electricity markets. Currently, she is a research engineer with the Energy Science & Technology Division, Pacific Northwest National Laboratory, Richland, WA. She was with Shenyang Electric Power Survey and Design Institute from 1993 to 1998.

**Z. Todd Taylor**

**David P. Chassin** (M'02) received his B.S. of Building Science from Rensselaer Polytechnic Institiute in Troy, New York. He is a staff scientist with the Energy Science & Technology Division at Pacific Northwest National Laboratory, where he has worked since 1992. He was Vice-President of Development for Image Systems Technology from 1987 to 1992, where he pioneered a hybrid raster/vector computer aided design (CAD) technology called CAD OverlayTM. He has experience in the development of building energy simulation and diagnostic systems, leading the development of Softdesk Energy and DOE's Whole Building Diagnostician. He has served on the International Alliance for Interoperability's Technical Advisory Group and chaired the Codes and Standards Group. His recent research focuses on emerging theories of complexity as they relate to high-performance simulation and modeling.

**Ross T. Guttromson** (M'01) received a B.S.E.E degree from Washington State University, Pullman. Currently, he is a senior research engineer with the Energy Science & Technology Division, Pacific Northwest National Laboratory, Richland, WA. He was with R. W. Beck Engineering and Consulting, Seattle, WA, from 1999 to 2001 and with the Generator Engineering Design Group, Siemens-Westinghouse Power Corporation, Orlando, FL, from 1995 to 1999. Ross has one U.S. and one international patent, and is author and co-author of several papers on power systems and distributed resources. He is a member of the WECC Load Modeling Task Force, IEEE SCC 21 P1547.2 standards group on Interconnecting Distributed Resources with Electric Power Systems, and the IEEE Wind Model Development Task Force. Ross is a U.S. Navy submarine veteran, having served on the USS Tautog (SSN 639) from 1987 to 1991.

**Scott Studham** received his B.S. (1997) and M.S. (2003) from Washington State University. He is currently managing the operations of the MSCF in the Environmental Molecular Sciences Laboratory (EMSL) at Pacific Northwest National Laboratory. His responsibilities include operations management of the Molecular Science Computing Facility, supercomputer upgrades, and supervision of EMSL's data management needs. He has a dual assignment to the Computational Sciences and Mathematics directorate, where he is responsible for High Performance Computing and Facilities Operations. From 1998 - 2000, Mr. Studham serviced the National Weather Service supercomputer center in Bowie MD where he led a team responsible for two supercomputers that are used to predict the nation's weather. From 1992 - 1998 he worked on consulting projects related to high performance computing. Previous to this he worked as a system administrator in support of computational chemistry for a research organization. Mr. Studham is an IBM Certified Advanced Technical Expert and IBM Certified Senior Program Manager.

Filename:              super_computer_internal_after dave Sue and Todd.doc
Directory:             C:\Documents and Settings\d3g637\Local Settings\Temporary Internet
                       Files\OLKF9
Template:              C:\Documents and Settings\d3g637\Application
                       Data\Microsoft\Templates\Normal.dot
Title:                 Parallel Computing Environments and Methods for Power Distribution
                       System Simulation
Subject:               IEEE Transactions on Power Systems
Author:                N. Lu, Z. T. Taylor, D. P. Chassin, R. T. Guttromson, R. S. Studham
Keywords:
Comments:              submitted 9/2004 to IEEE Transactions on Power Systems
Creation Date:         9/9/2004 4:23 PM
Change Number:         2
Last Saved On:         9/9/2004 4:23 PM
Last Saved By:         Ning Lu
Total Editing Time:    4 Minutes
Last Printed On:       9/18/2004 9:03 AM
As of Last Complete Printing
      Number of Pages:        6
      Number of Words:        3,734 (approx.)
      Number of Characters:   21,287 (approx.)